\documentclass[aps,prl,superscriptaddress,groupedaddress]{revtex4}
\usepackage[toc,page]{appendix} 
\usepackage{amsmath,amssymb,graphics,setspace}
\usepackage{graphicx}
\usepackage{dcolumn}
\usepackage{bm}
\usepackage{subfigure}
\usepackage{color}
\newcommand{\mathsym}[1]{{}}
\newcommand{\unicode}[1]{{}}
\newcommand{\ber}{\begin{eqnarray}}
\newcommand{\eer}{\end{eqnarray}}
\newcommand{\bea}{\begin{equation}}
\newcommand{\eea}{\end{equation}}

\begin{document}

\title{Galaxy rotation curves from external influence on Schwarzschild geometry}

\author{A. Bhattacharyay}
\affiliation{Indian Institute of Science Education and Research, Pune, India}

\begin{abstract}
We present a modified Schwarzschild metric to introduce a weak breakdown of asymptotic flatness. The kinematics of the metric captures a wide range of galaxy rotation curves. We show baryonic Tully-Fisher relation on the basis of this modified Schwarzschild metric. On the basis of the kinematics of this modified metric we also show some connections between the size and ordinary matter content of the observable universe and the rotation curves of spiral galaxies.
\end{abstract}
\maketitle
\section{Introduction}
Galaxy rotation curves of spiral (disk) galaxies show that the magnitude of velocity $v$ of distant stars is, to a good extent, independent of the distance $r$ from the center of the galaxy. This is a regime supposed to be of Newtonian gravity where rotation curves exhibit $r$ independence of $v$. According to Newton's law, the velocity of approximate mean circular motion of these stars should follow the relation
\bea
\frac{v^2}{r} = \frac{\text {G}M}{r^2},
\eea 
where $M$ is the luminous mass at the galaxy center. This gives $v = \sqrt{\text {G}M/r}$ and it should fall with distance as $r^{-\frac{1}{2}}$, however, as observations show \cite{gal,rub}, it remains relatively constant or even increasing with distance in some cases.
 
Existence of dark matter is invoked in this scenario (and in many other cases, e.g., gravitational lensing) where one considers the mass of the galaxy comprises of an otherwise invisible component which is spread over being proportional to the distance i.e., $M(r)\propto r$ \cite{roos}. This can produce velocity profiles as seen at large $r$ to be independent of $r$. This idea of existence of dark matter is probably the most dominant paradigm followed at present.

People have tried to explain these rotation curves by invoking different ideas within the realm of Einstein's gravity. An interesting idea in this respect is the one put forward by Carmeli \cite{carm} couple of decades ago. Carmeli tried to take into account the effects of the expanding universe on the rotation curves at large distances. More work has been done along this line following Carmeli's 1998 work \cite{hart}. There are other recent approaches to explain galaxy rotation curves using conformal gravity \cite{mann,li}, Palatini formalism \cite{spor}, using Grumiller's modified gravity \cite{lin} etc. Efforts have also been made to show dark matter as a cosmological relativistic gravitational effect and not arising from exotic particles which are, in general, outside the scope of standard model \cite{kra,kra1,bole,bol1,roy,mus,buch,qui,boud}.

Another competing idea to explain galaxy rotation curves is due to Milgrom who considered altering Newton's law of motion or equivalently gravity when the acceleration is extremely small of the order of $10^{-8}$ cm/$\text s^2$ \cite{mil1,mil2,mil3,mil4,mil5,san}. This is an interesting idea applicable to galaxy rotation curves and is followed by many to explain such observations \cite{acid,scar}. This paradigm of MOND does not really justify (within the scope of known facts) why Newton's law of motion should alter, however, accepts the observed facts that it alters in the case of such galaxy rotation curves. This is an approach very different from consideration of dark matter, however, the MOND paradigm is consistent with the idea of asymptotic flatness. In the present paper, we propose a {\it purely kinematical examination} of a model of modified Schwarzschild metric where asymptotic flatness is weakly broken.

The non-Newtonian nature of gravity at the edge of the spiral galaxies as shown by rotation curves indicates a possibility of non existence of asymptotic flatness for spiral galaxies. This is the main hypothesis of this paper. On the basis of this hypothesis, we explore here the modification of the Newtonian regime of Schwarzschild gravity in order to demonstrate the possibility of existence of galaxy rotation curves. We show in this paper that the galaxy rotation curves over a good range of variety can be captured by the proposed modified Schwarzschild metric which deviates from asymptotic flatness in a particular way. Our phenomenology correctly derives the baryonic Tully-Fisher relation with the observed order of magnitude of its amplitude. As an offshoot, the present analysis shows some interesting connections between the orders of magnitude of the radius and the mass of ordinary matter of the observable universe and the parameter of the modified metric showing rotation curves. 

Note that, McGaugh et al., have shown correlations between observed and baryonic components of accelerations considering a wide range of galaxies in ref.\cite{mc1,mc2}. In these papers, {\it each and every plot of observed acceleration against the baryoinic acceleration shows tendency of asymptotic saturation of the observed acceleration where the baryonic one goes to zero}. Note also that, there are some issues with respect to the possible best method of sorting through the lowest acceleration data which have been discussed at length in ref.\cite{mc3,mc4}. In view of these issues, the fitting functions for such plots have been chosen in a way to maintain the constancy of velocity in the rotation curves at smaller accelerations. Considering this observed tendency of the acceleration to saturate at small values, in our present work, we would investigate the situation where the asymptotic flatness is broken by the existence of a constant acceleration of the value of that in the MOND regime. In the weak fields (MOND) regime, Newton's force of gravity would be comparable to this force and at even larger distances this constant force per unit mass will dominate until it is altered by the presence of another galaxy. 

Asymptotic flatness is an idealization to implement isolation of a local system from the rest of the observable universe such that conservations can be imposed in the absence of gravitational surroundings. In this context, the article by Roberts \cite{rob} has interesting discussion. Galaxy rotation curves are indication of clear departure from Newtonian gravity is the central idea on which we rely in the present analysis. In what follows, we will guess the modified Schwarzschild metric by the demand that it has to add a constant acceleration to that of Newtonian gravity. The modification of the Schwarzschild metric is assumed to be due to sources external to a galaxy. We try to justify this assumption {\it a posteriori} by indicating some very interesting connections of orders of magnitute of numbers at the scales of observable universe and those of spiral galaxies.

The paper is organized in the following way. We first present the modified Schwarzschild metric and its immediate kinematic consequences. In this part we show the derivation of the Tully-Fisher relation on the basis of the modified metric and also establish a connection between acceleration scale of the MOND regime and the order of magnitude of the radius of the observable universe. Then we explore the energy-momentum tensor corresponding to the considered metric to identify the pressure and density of the field that could be associated to the origin of this metric. We show that the field can have its pressure proportional to its density. We also show that the density when integrated over the region of the observable universe results in the right orders of magnitude of the baryonic matter content of the observable universe. We conclude the paper by a discussion on the important results of the present phenomenology.

\section{Modified Schwarzschild Metric}
Consider the modification of the Schwarzschild metric which would add a constant acceleration to the Newtonian gravity at large distances. This constant background acceleration is expected to come out to be of the order of $10^{-10}$ m/$\text s^2$ (MOND regime) such that it only affects weak Newtonian gravity. A metric that can result in such a uniform background  central force field in the presence of a distribution of matter on top of the Schwarzschild mass parameter can be written as 
\bea
ds^2=(1-\frac{\lambda}{r}+\beta r)c^2dt^2-\frac{1}{1-\frac{\lambda}{r}-\delta(\beta r)}dr^2 - r^2(d{\theta}^2+\sin^2{\theta}d{\phi}^2),
\eea
where $\beta$ is an inverse length scale to be found out from the parameters of a typical disk galaxy, $\delta$ is a dimensionless positive parameter to be found out as well and $\lambda$ is the Schwarzschild radius $2\text G M/c^2$. The inclusion of the terms proportional to $\beta r$ when $\beta$ is very small will weakly break the asymptotic flatness of the Schwarzschile metric resulting in a constant inward acceleration which is comparable to the acceleration due to Newtonian gravity in the MOND regime.  

Let us have a look at the nonzero elements (diagonal) of the Ricci tensor of this metric. These elements are
\ber\nonumber
R_{00} &=& -\frac{c^2\beta\left ( 4r^4\beta^2\delta +r^3\beta(5\delta -3)+r(\delta +9)\lambda -2(\delta+3)\lambda^2 + r^2(\beta\lambda(2-6\delta)-4) \right )}{4r^2(r+r^2\beta-\lambda)},\\\nonumber
R_{11} &=& -\frac{\beta\left ( 4r^4\beta^2\delta +r^3\beta(1+9\delta) - r(3+11\delta)\lambda +2(1+3\delta)\lambda^2 - 2r^2(3\beta\lambda +\delta(7\beta\lambda -2)) \right )}{4(r+r^2\beta-\lambda)^2(-r+r^2\beta\delta+\lambda)},\\\nonumber
R_{22} &=& \frac{r\beta\left ( 4r^2\beta\delta +r(3\delta -1)-2(\delta -1)\lambda \right )}{2(r+r^2\beta -\lambda)},\\\nonumber
R_{33} &=& \frac{r\beta\left ( 4r^2\beta\delta +r(3\delta -1)-2(\delta -1)\lambda \right )\sin^2{\theta}}{2(r+r^2\beta -\lambda)},
\eer
where 0, 1, 2, and 3 in the indices correspond to $t$, $r$, $\theta$ and $\phi$. In the following, it will be shown that for the galaxy rotation curves to appear at distance $r$, $\beta r^2 \simeq \lambda$. This relation $\beta r^2 \simeq \lambda$ will turn out, in what follows, to be the central result of this paper and has to be verified by observations. 

The expressions of the nonzero Ricci tensor elements indicate $R_{22}, R_{33} \sim 1/r$ (and $R_{00}$, $R_{11}$ are even smaller). It can also be checked that all the elements of the Riemann tensor are devoid of any divergence at large $r$. The only singularity is at $r=0$ which is standard. As we will see in the following, the $\beta r$ perturbation will come out to be order unity at the edge of the observable universe, hence, there is no large $r$ issue at length scales smaller than that. 

The weak breaking of the limit of asymptotic flatness of the Schwarzschild metric, in the present scenario, is never a problem at larger distances. At large distances from a galaxy, in reality, the spacetime is really not flat due to the presence of other sources (e.g., galaxies). The consideration of asymptotic flatness helps keep this reality aside to treat a galaxy in an isolated manner. What the present modified Schwarzschild metric does is that it alters the idealized asymptotic flatness condition very weakly.

Consider the Geodesic equations of this system as are shown below where $\tau$ is an affine parameter -
\ber\nonumber
&& \frac{d^2r}{d\tau^2}+\frac{-r^2\beta\delta+\lambda}{2r(-r+r^2\beta\delta+\lambda)}\left (\frac{dr}{d\tau}\right )^2-(r-r^2\beta\delta -\lambda)\left (\frac{d\theta}{d\tau}\right )^2\\ &&-(r-r^2\beta\delta -\lambda)\sin^2{\theta}\left (\frac{d\phi}{d\tau}\right )^2+\frac{c^2}{2r}\left ( \beta +\frac{\lambda}{r^2} \right)\left ( r-r^2\beta\delta -\lambda \right )\left (\frac{dt}{d\tau} \right )^2=0\\
&& \frac{d^2\theta}{d\tau^2} +\frac{2}{r}\frac{dr}{d\tau}\frac{d\theta}{d\tau} - \cos{\theta}\sin{\theta}\left (\frac{d\phi}{d\tau} \right )^2=0\\
&& \frac{d^2\phi}{d\tau^2}+\frac{2}{r}\frac{dr}{d\tau}\frac{d\phi}{d\tau}+2\cot{\theta}\frac{d\theta}{d\tau}\frac{d\phi}{d\tau}=0\\
&& \frac{dt}{d\tau^2}+\frac{r\left ( \beta +\frac{\lambda}{r^2} \right )}{r+r^2\beta -\lambda}\frac{dr}{d\tau}\frac{dt}{d\tau}=0.
\eer

The case of a circular orbit (for the sake of simplicity) on the equatorial plane $\theta = \pi/2$ makes $t \propto \tau$ and from eqn.(3)
\bea
v = \frac{c}{\sqrt{2}}\sqrt{\left (\beta r + \frac{\lambda}{r}\right )}
\eea
where on this circular orbit $v = r\frac{d\phi}{dt}$ and $\frac{d^2r}{d\tau^2}=\frac{dr}{d\tau}=\frac{d^2\theta}{d\tau^2}=\frac{d\theta}{d\tau}=\frac{d^2\phi}{d\tau^2}=0$. The eqn.(7) immediately indicates that $\beta r$ has to be comparable and within an order of magnitude of $\lambda/r$ to effectively raise the rotation curves above what results from the Newtonian gravity.

The expression of the velocity reveals that if $\beta$ is too small then for smaller $r$ and corresponding $\lambda$ there is no reason not to get Keplerian orbits which are hardly perturbed. For example, this is the case of our solar system. The Newtonian gravity, as obtained from perturbation of Minkowski space would remain practically intact for the inner part of planetary systems for a very small $\beta$ despite the fact that orbits at the edge of disk galaxies can get altered from what Newtonian gravity predicts. The $\beta$ must be very small for this phenomenology to work and, using some known numbers, let us have an estimate of how small the $\beta$ is. 

Take the example of our galaxy the Milky Way. It has a central star Sagittarius A* (considered to be a super massive black hole) of radius of the order of $10^7$ km and a mass of about 4 million solar masses. The luminous mass of the same galaxy is estimated to be that of about 200 billion stars. Therefore, the $\lambda$ for such a system can be roughly estimated to be $10^{10}$ km considering the mass of the core of the galaxy to be about 4 billion solar masses. Now, 1 kpc being $10^{16} $ km, the distance $r$ in the scale of $\lambda$ is of the order of $10^6\times n$ where $n$ is a number that is safely considered to be within order 10. This immediately tells us that $r_0 = \beta^{-1}$ is of the order of $10^{13}$ in the units of $\lambda$. 1 kpc being $10^6$ in $\lambda$, $r_0 = 10^7 $ kpc. The radius of the observable universe being about $14.3 \times 10^9$ parsec is of the order of $r_0$.

It is now obvious, from analogy of Newtonian gravity, that the perturbation $\beta r$ in $g_{00}$ results in an uniform inward radial acceleration $|a_0| \simeq \nabla (c^2\beta r) = \beta c^2$ on top of what is there due to the Newtonian gravity as would be seen by a test particle. Knowing the estimation of $\beta = 1/10^{26}$ $\text m^{-1}$, one gets $a_0 \sim 10^{-10}$ $\text m/\text s^2$. This is exactly the force per unit mass regime of the MOND where the rotation curves are observed. As seen from the galaxy center, the Newtonian gravitational force will fall as $1/r^2$ and a distance independent central force per unit mass $\vec{a}_0$ will be seen to exist over large distances where the principle of superposition holds.

Let us quickly check that equation (7) can account for the baryonic Tully-Fisher relation \cite{lelli,mcga,tully} where the mass of a spiral galaxy is proportional to the forth power of the distance independent velocity of stars in the rotation curve. Considering $\beta r \simeq \lambda/r$, $r \simeq \sqrt{\lambda/\beta}$. This gives $v \simeq c(\lambda\beta)^{\frac{1}{4}}$, which will result in 
\bea\nonumber
M \simeq \left (\frac{1}{2\beta\text G c^2}\right )v^4,
\eea
where we have used the Schwarzschild radius $\lambda=2\text G M/c^2$ and $M$ is the baryonic mass responsible for the Schwarzschild geometry considered. One particular thing to notice about the constant of proportionality is that, since $\beta c^2 = 9\times10^{-10}$ m/$\text s^2$ and $\text G=6.6\times 10^{-11}$ $ \text m^3$/(kg $\text s^2$), $1/(2\beta\text G c^2)$ is about $10^{20}$ kg $\text s^4/\text m^4$. The observed normalization factor $A$ of barynic Tully-Fisher relation is ${A} = 50$ $\text M_{\odot}$ $\text s^4/\text {km}^4$ which, considering $M_\odot = 2 \times 10^{30}$ kg, is about $10^{20}$ kg $\text s^4$/$\text m^4$ \cite{mcga}. The observed number and the calculated one being remarkably close the universality of $\beta$ can be inferred. Note that, the magnitude of $\beta$ we have fixed here quite heuristically by considering the mass of the core of the Milky Way to be a billion solar mass. However, if the same has to work for all galaxies, $\beta$ has to be universal and matching the baryonic Tully-Fisher relation lends some support.

If $r_0=\beta^{-1}$ is universal, then for all spiral galaxies with non Newtonian rotation curves, the radius of the galaxy edge could be estimated as $\sqrt{\lambda r_0}$ where $\lambda$ is the Schwarzschild radius of the galaxy. Note that, in the present phenomenology, the $\lambda$ of a galaxy must be estimated on the basis of its baryonic mass and no dark matter is involved. Like the baryonic Tully-Fisher relation, there must exist another power law relation between the mass giving the Schwarzschild radius and the radius of such galaxies. The relation is given by $M = (c^2/2\text G r_0)\times r^2$ where the amplitude $c^2/2\text G r_0$ can be readily estimated to be of the order of 10 kg/$\text m^2$ which, apart from the power law, could also be tested by observations. 

\begin{figure}[h]
  \includegraphics[width=8cm]{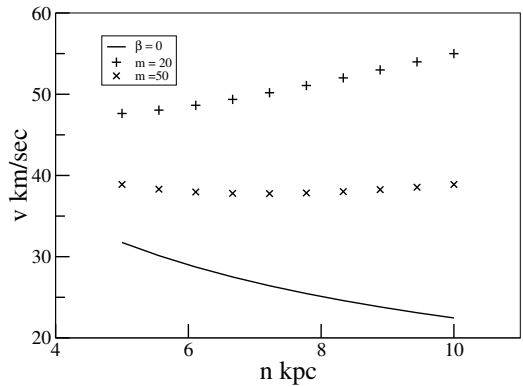}
  \caption{\small {Velocity vs distance (parametrized by $n$) plot shows that depending upon the parameter $m$ which could be a number of order 10. The plot shows existence of a wide range of rotation curves with velocities bigger in magnitude than that given by the Newtonian gravity at these distances.}}
\label{fig:Local diffusivity}
\end{figure}

Note that, in none of the above estimation dark matter contribution is taken into account which actually makes the mass of our galaxy to be 1.5 trillion solar masses. The important lesson this order of magnitude calculations give us is that, if the size of the galaxy scaled by its mass has some universality then that probably is due to the rest of the observable universe the galaxy is in causal connection with. This connection throws up the correct order of magnitude of $r_0$ based only on local information of a spiral galaxy. Let's get the velocity $v$ in km/sec using these numbers. 
\bea
v = \frac{c}{\sqrt{2}}\sqrt{\left (\frac{r}{r_0} + \frac{1}{r}\right )} = \frac{10^5}{\sqrt{2}\times 10^3}\sqrt{\left (\frac{n}{10} + \frac{1}{n}\right )} = \frac{100}{\sqrt{2}}\sqrt{\left (\frac{n}{10} + \frac{1}{n}\right )},
\eea
where $r_0=1/\beta\lambda$ when all the distances are scaled by $\lambda$. It can immediately be recognized that we are getting the right orders of magnitude for the velocity. For a visual representation of the effect refer to the Fig.1 where the velocity 
\bea
v = \frac{100}{\sqrt{2}}\sqrt{\left (\frac{n}{m} + \frac{1}{n}\right )}
\eea
has been plotted for two different choices of the parameter $m$ which can safely be considered to be a number of the order of a few 10. These curves are compared with the unperturbed (Newtonian) one corresponding to $\beta = 0$. The number $n$ which parametrizes distance, according to our present estimation should be within order 10. 

The present modification of Schwarzschild metric would not affect smaller length scales (for example solar system) due to the very large value of the parameter $r_0$. Regions where $\lambda/r^2$ is considerably bigger than the $1/r_0$, the Newtonian limit of gravity will be the dominant contributor. This is the regime of the planetary systems which show Keplerian orbits. For example, the Schwarzschild radius of the Sun being about 3 km, the value of $r$ at which $\lambda/r^2$ is comparable with beta is $r \sim 10^{12}$ km. The average distance between the Sun and the Pluto is only about $6\times 10^{9}$ km which is much less than $10^{12}$ km. Therefore, one can safely go by the consideration of asymptotic flatness for such planetary systems. 

\section{Einstein Tensor of the modified Schwarzschild metric}
The elements of the Einstein tensor corresponding to the metric given in eqn.(2) are 
\ber
\text G_{00}&=& \frac{2c^2\beta\delta(r+r^2\beta-\lambda)}{r^2}\\ 
\text G_{11} &=& \frac{\beta(r(\delta -1)+2r^2\beta\delta+2\lambda)}{(r+r^2\beta -\lambda)(-r+r^2\beta\delta+\lambda)}\\
\text G_{22}&=&-\frac{r\beta\left ( 4r^4\beta^2\delta +r^3\beta(7\delta-1)+r(3-5\delta)\lambda +2(\delta -1)\lambda^2-2r^2(1+\beta\lambda +\delta(5\beta\lambda -1)) \right )}{4(r+r^2\beta -\lambda)^2}\\
\text G_{33}&=&\sin^2{\theta}\times\text G_{22}
\eer
Note that, if $\delta=0$ there is no density to account for the pressure which is proportional to the parameter $\beta$. Considering that $\beta r << 1$ even at the scales where the rotation curves are observed where $r >> \lambda $ the leading order contributions to the elements of the Einstein tensor will come out to be
\ber
\text G_{00}&=& \frac{2c^2\beta\delta}{r}\\ 
\text G_{11} &=& \frac{\beta}{r}(1-\delta)\\
\text G_{22} &=& \frac{\beta r(1-\delta)}{2}\\
\text G_{33}&=&\sin^2{\theta}\times\text G_{22}.
\eer
The leading order expressions indicates that for the pressure to remain positive, $0<\delta < 1$. In this order, one may safely consider $\delta \sim 1$ for the sake of simplicity. The density and pressure, corresponding to the fields that results in the modified metric, are proportional to $\beta/r$. This proportionality indicates that the source could be some gas or even radiation. 

From the above mentioned relation (14) we get a mass density $\rho \simeq \frac{c^2 \beta}{4\pi \text G r}$ where $\delta \sim 1$. Let us see, when integrated over the radius of the observable universe (radius $r_0 = \beta^{-1}$), what this density $\rho$ results in. This results in the right order of magnitude of the estimated mass of the ordinary matter of the observable universe $M_{uni} = \frac{c^2\beta r_0^2}{2\text G}\simeq \frac{c^2 r_0}{2\text G} $ since $\beta \sim r_0^{-1}$. Putting the typical values, $c^2 = 9\times 10^{16}$ $ \text m^2/\text s^2$, $r_0 = 10^7\times 10^{19} $ m and $\text G=6.6\times 10^{-11}$ $ \text m^3$/(kg $\text s^2$), the mass of the ordinary matter of the observable universe comes out to be $M_{uni} \simeq 10^{53}$ kg.  

So, we get an expression of the ordinary matter of the observable universe in terms of the universal constants $c$, G and the radius of the observable universe $r_0$ in connection with the modified Schwarzschild geometry which is capable of capturing rotation curves. The Schwarzschild metric is modified in a way such that on top of the Schwarzschild mass parameter there exists a distribution of matter density falling as $\beta/r$. With this modification, the small $r$ character of the metric remains practically Schwarzschild whereas the region at large distances gets modified. This connection possibly indicates that the weak loss of asymptotic flatness of local spacetime of a spiral galaxy may have some relation with the surroundings of the galaxy which is the observable universe.

Interestingly, the importance of consideration of the mass of gas content of such a galaxy for baryonic Tully-Fisher relation is highlighted a couple of decades ago in the paper of McGaugh et al., in ref.\cite{mcga}. One can simply consider the perfect fluid to be a gas and the adjustment of the parameter $\delta$ will depend on the details of the equation of state of the gas. The constant parameter $\lambda$, which in the present context is the standard Schwarzschild parameter, does not contribute to anything in the energy-momentum tensor just like what happens in the case of vacuum Schwarzschild solution.

\section{Discussion}
Based on the idea that galaxy rotation curves indicate a weak breaking of asymptotic flatness which is considered to isolate a galaxy from its surroundings, we have proposed a modified Schwarzschild metric. We have done a purely kinematical analysis of this modified metric. This idea of weak breakdown of asymptotic flatness also has some observational support coming from the reported observations of McGaugh et al., \cite{mc1,mc2}. We show in this paper that a wide range of galaxy rotation curves can be captured by the modified metric.

All the results of this paper come from the consideration of a phenomenological length scale $r_0$ of the order of the radius of the observable universe. In the present phenomenology, this length scale bears a very close relation with the magnitude of acceleration in the MOND regime. We have shown that, the mass density and pressure corresponding to the modified metric are proportional and scale as $1/r$. An integration of this mass density over the volume of the observable universe correctly gives the estimated order of magnitude of ordinary matter of the observable universe. This is some indication that the non existence of the asymptotic flatness of the metric considered could be some gross effect of its surroundings.

Our simple model of breaking the asymptotic flatness by the presence of a constant acceleration captures the baryonic Tully-Fisher relation not only in the correct power law, but also in the correct estimation of the amplitude (proportionality constant). This in fact is a remarkable connection which is produced by the present phenomenological model besides capturing the correct order of magnitude of acceleration of the MOND regime in relation with the radius of the observable universe.

\section{Acknowledgement}
I acknowledge very helpful discussions with Stacy S McGaugh on baryonic Tully-Fisher relation and correlation of accelerations.

\end{document}